\documentclass[11pt,twoside]{article}
\usepackage{latex8}
\usepackage[dvips]{epsfig}
\usepackage[dvips]{graphicx}

\usepackage[latin9]{inputenc} \usepackage[T1]{fontenc}
\usepackage[english]{babel}

\usepackage{amsmath}
\usepackage{amsfonts}
\usepackage{amssymb,amsthm}
\usepackage{fancybox}
\usepackage{url}
\usepackage{graphicx}
\usepackage{xypic}
\usepackage{multicol}
\usepackage{rotating}
\usepackage{listings}
\usepackage{algorithm,color}

\newcommand{\old}[1]{\mathrm{\backslash{}old}\left(#1\right)}

\begin{document}

\title{\bf Formal proof for delayed finite field arithmetic \\ using
  floating point operators\thanks{This work has been partially founded
    by PICS 2533 of the CNRS, project EVA-Flo of the ANR, project
    CerPAN of the ANR and PPF Suréna.}}

\author{%
  Sylvie Boldo  \\ {\sc inria}  (Saclay - Île-de-France)               \\ sylvie.boldo@inria.fr    \and
  Marc Daumas   \\ {\sc eliaus} ({\sc ea 3679 upvd})                   \\ marc.daumas@ens-lyon.org \and
  Pascal Giorgi \\ {\sc lirmm}  ({\sc umr} 5506 {\sc cnrs}--{\sc um}2) \\ pascal.giorgi@lirmm.fr
}

\maketitle \thispagestyle{empty}
  
\begin{abstract}
  Formal proof checkers such as Coq are capable of validating proofs
  of correction of algorithms for finite field arithmetics but they
  require extensive training from potential users.  The delayed
  solution of a triangular system over a finite field mixes operations
  on integers and operations on floating point numbers. We focus in
  this report on verifying proof obligations that state that no round
  off error occurred on any of the floating point operations. We use a
  tool named Gappa that can be learned in a matter of minutes to
  generate proofs related to floating point arithmetic and hide
  technicalities of formal proof checkers. We found that three
  facilities are missing from existing tools.  The first one is the
  ability to use in Gappa new lemmas that cannot be easily expressed
  as rewriting rules. We coined the second one ``variable
  interchange'' as it would be required to validate loop interchanges.
  The third facility handles massive loop unrolling and argument
  instantiation by generating traces of execution for a large number
  of cases. We hope that these facilities may sometime in the future
  be integrated into mainstream code validation.
\end{abstract}

\newcommand{\R}{\mathbb{R}}
\newcommand{\Z}{\mathbb{Z}}
\newcommand{\IF}{\mathbb{IF}}
\newcommand{\Zp}{{\Z/p\Z}}

\lstset{basicstyle=\tt\small,xleftmargin=0cm,numberstyle=\tiny}

\renewcommand{\thelstlisting}{\arabic{lstlisting}}

\section{Introduction}

Introducing a new algorithm is a difficult task. Authors have to
persuade readers that their algorithm is correct and efficient. Such 
goals  are usually attained  by providing pen-and-paper proofs of
correction more or less interlaced with  the description of the algorithm.
Authors may also provide results of tests to guarantee correction
and efficiency on random  cases and on known or new hard
cases.  Alas, this process is known to fail on mundane  as well as notorious
occurrences~\cite{RusHen91}.

Developing a proof of correction in a formal proof checker using
higher order logic such as Coq~\cite{BerCas04} would be a nice
alternative but such a task usually represents a large amount of work
outside the fields of expertise of most authors.

The delayed solver studied here works on a $N \times N$ unitary
triangular matrix on $\Z / p \Z$ finite field. The key improvement of
this algorithm compared to state of the art lies in the fact that
delayed algorithms use floating point units to perform operations with
no rounding error and delay computations of remainders as much as
possible. Operations on floating point numbers are limited to three
functions. The other functions use combinatorial logic.

The first function (\verb+DGEMM_NEG+) performs a naive matrix
multiplication and Gappa may soon be able to handle the proof
obligation generated by a tool such as the Why
platform~\cite{FilMar07} and the corresponding floating-point
annotations~\cite{BolFil07}.  The second function (\verb+DTRSM+) is
invoked only under the \verb+delay+ predicate.  This is enforced by
the condition on the induction of the invoking function
(\verb+LZ_TRSM+) for $N$ between 2 and 54 with $p$ prime varying
between $2$ and $94,906,266$. Variable interchange in the \verb+delay+
predicate allows to limit the proof to the 53 different values of $N$
where a naive user would consider the $94,906,265$ different values of
$p$.

Proof obligations are usually derived from a static analysis of the
source code considered. Our work showed that generating proof
obligations from traces of execution after most parameters have been
instantiated may also be useful. We have set up a C++ class to provide
such proof obligations but we hope that such capability will be
provided by Why and similar tools in the future.

We present some background information in the remaining of the
introduction. We continue with a new lemma that might be used by Gappa
for inductive linear bounds in Section~\ref{sec/had} and with our
prototyping variable interchange developments in
Sections~\ref{sec/xch}. We conclude this work in
Section~\ref{sec/conc}.

\newcommand{\lCeil}{\left\lceil}
\newcommand{\rCeil}{\right\rceil}
\def\so{{O {\;\!\tilde{}}\,}}
\newcommand{\linbox}{{\sc LinBox}}
\newcommand{\lsp}{{{\tt LSP}}}
\newcommand{\lup}{{{\tt LUP}}}
\newcommand{\lud}{{{\tt LUdivine}}}
\newcommand{\lqup}{{{\tt LQUP}}}
\newcommand{\turbo}{{{\tt TURBO}}}
\newcommand{\fgemm}{{{\tt Fgemm}}}
\newcommand{\tu}{{{\tt TURBO}}}
\newcommand{\trsm}{{{\tt Trsm}}}
\newcommand{\ltrsm}{{{\tt ULeft-Trsm}}}
\newcommand{\ultrsm}{{{\tt ULeft-Trsm}}}
\newcommand{\lltrsm}{{{\tt LLeft-Trsm}}}
\newcommand{\urtrsm}{{{\tt URight-Trsm}}}
\newcommand{\lrtrsm}{{{\tt LRight-Trsm}}}
\newcommand{\dbl}{\texttt{double} }

\newcommand{\di}{\displaystyle}
\newcommand{\GF}[1]{\ensuremath{\mathtt {GF}(#1)}}
\newcommand{\GO}{{\cal O}}
\newcommand{\pF}[1]{\leavevmode
        \kern.1em\raise.0ex \hbox{\Z}\kern-.1em /\kern-.15em\lower.3ex
         \hbox{#1}\mbox{\Z}}
\newcommand{\til}{\lower 2pt\hbox{\small${}^\sim$}}

\newtheorem{thm}{Theorem}[section]
\newtheorem{cnj}[thm]{Conjecture}
\newtheorem{lem}[thm]{Lemma}
\newtheorem*{cor}{Corollary}
\newtheorem{prop}[thm]{Proposition}
\newtheorem{defi}[thm]{Definition}
\newtheorem{rem}[thm]{Remark}
\newtheorem{pty}[thm]{Property}

\subsection{Finite field arithmetic and application to linear algebra}


Finite field arithmetic plays a crucial role in nowadays applications.
One of the most extensively studied application of finite fields is
cryptography.  Another key application of finite field arithmetic
arises with exact linear algebra computation where modular techniques
(e.g. CRT or P-adic lifting) allow some control on expression swell
with high performances (see~\cite{EGGSV06-1} and references herein).
While cryptographic applications need finite fields of large
cardinality for security purpose, most exact linear algebra restrains
to machine word size prime field (e.g. 32 or 64 bits) in order to
benefit from machine arithmetic units.

A classical way to perform one arithmetic operation in a prime field, here we refer to integers modulo a prime number,
is to first perform the operation on integers and second reduce the result to the destination field.
Let  $x,y\in \Zp$ and $* \in \{ +, \times \}$. One may compute $z=x*y\in\Zp$ by computing $t=x*y \in \Z$ and a modular reduction $z = t \mod p$.




When one deals with fixed precision prime field arithmetic, two majors
issues arise:\, performances and cardinality limitation. The latter
issue can have a non-negligible impact on the former one. As was just
said, the classical way to perform arithmetic operations over a prime
field is to perform operations on integers and reduce intermediate
results.  Therefore, all integers between $0$ and $(p-1)^2$ must be
representable to correctly perform multiplications over $\Zp$. This
limitation slightly increase to perform an \texttt{AXPY} operation (a
multiplication followed by an addition) with only one reduction step.
This implies that all integers between $0$ and $p\times(p-1)$ must be
representable.



Using word-size machine integers and classic arithmetic we obtain the
following cardinality limitation: $p<2^{16}$ on 32 bit architectures
and $p<2^{32}$ on 64 bit architectures with {\tt unsigned} types.  An
alternative to increase cardinality of word-size prime fields is to
use floating point numbers.  According to the IEEE~754
standard~\cite{Gol91}, mantissas of double precision floating point
numbers can store 53 bit integers (including the implicit bit).
Therefore, we can perform prime field arithmetic with cardinality up
to $2^{26}$ using \texttt{double}. Note that, the reduction is easily
obtained by the \texttt{fmod} function available in standard
libraries.  This approach is quite interesting in practice since
floating point multiplications and divisions may be faster than their
integer counterparts.



On selected classes of algorithms, delayed prime field arithmetic
sustains better performances.  The idea is to perform several integer
operations before reduction into the field.  It has been very fruitful
for exact linear algebra~\cite{jgd:2004:issac}.  Delayed exact linear
algebra computations also benefit from optimized numerical BLAS (e.g.
ATLAS~\cite{Whaley:2001:AEO}, GOTO~\cite{2002:gotoblas}) libraries for
exact computations and they often reach peak FPU throughput for
operations over a finite field.

Beside basics linear algebra operations such as matrix-vector products
and matrix multiplications, delayed arithmetic over a prime field is
valuable when expressions swell largely such as solving systems of
linear equations.  This approach works perfectly for unitary
triangular system (only ones along the diagonal) despite the
exponential growth of the intermediate variables.


\subsection{Formal proof checking and Gappa}

Gappa~\cite{DauMel04} has been created to generate formal certificates
of correction for programs that use floating point
arithmetic~\cite{DinLauMel06,MicTisVey06,MelPio07} and is related to
other developments~\cite{Har2Ka,DauMelMun05}. It is available from
\begin{center}
  \url{http://lipforge.ens-lyon.fr/www/gappa/}.
\end{center}
It will in the future be able to interact seamlessly with
Why~\cite{BolFil07}, a tool to certify programs written in a generic
language. C and Java can be converted to this language.

Gappa manipulates arithmetic expressions on real and rational numbers
and their evaluations on computers. Exact and rounded expressions are
bounded using interval arithmetic~\cite{JauKieDidWal01}, forward error
analysis and properties of dyadic fractions.  To the authors' best
knowledge, Gappa is the first tool that can convert some of the simple
tasks performed here into formal proofs validated by an automatic
proof checker.  Such goal has previously been quoted as {\em invisible
  formal methods}~\cite{TiwShaRus03} in the sense that Gappa delivers
formal certificates to users that are not expected to write any piece
of proof in any formal proof system.


% ...

Gappa produces a Coq file for a given input script.  Users do not need
to be able to write the Coq file but they can check the work of Gappa
by reading it or parsing it automatically.  It contains {\tt
  Variable}s, {\tt Defini\-tion}s, {\tt Notation}s, {\tt Lemma}s and
comments are between {\tt (*} and {\tt *)} signs.
Although enclosure is the only predicate available to users, Gappa
internally relies on more predicates to describe properties on
expressions.  All the properties of the input script are defined in
the Coq file.  Validity of proofs can automatically be checked by Coq.
More insights to Gappa are presented in~\cite{DauMel07}.

\section{A new lemma that might be used by Gappa for inductive linear
  bounds}
\label{sec/had}

\begin{figure}
\hrule
\smallskip
{\bf Input: } $A \in \Zp^{N \times N}$, $B \in \Zp^{N \times K}$.\\
{\bf Output: } $X \in \Zp^{N \times K}$ such that $AX=B$.
\smallskip
\hrule
\begin{quote}
{\bf if} N=1 {\bf then}\\
\hspace*{1cm} $ X:= A_{1,1}^{-1} \times B$.\\
{\bf else}
{\it (splitting matrices into $\lfloor \frac{N}{2} \rfloor$ and $\lceil \frac{N}{2} \rceil$ blocks) }\\[-.2cm]
\[
\begin{array}{cccc}
A & X & & B \\
\overbrace {\left[ \begin{array}{cc} A_1 & A_2 \\ & A_3 \end{array} \right] }&
\overbrace{\left[ \begin{array}{ccc} & X_1 & \\ & X_2 & \end{array} \right] }&
= &
\overbrace{\left[ \begin{array}{ccc} & B_1 & \\ & B_2 & \end{array} \right] }
\end{array}
\]
\hspace*{1cm}$X_2:=${\tt LZ\_TRSM($A_3,B_2$)}. \\
\hspace*{1cm}$B_1:= B_1 - A_2X_2$. \\
\hspace*{1cm}$X_1:=${\tt LZ\_TRSM($A_1,B_1$)}. \\
{\bf return } X.
\end{quote}
\hrule
\caption{First algorithm for {\tt LZ\_TRSM($A,B$)}}
\label{trsm}
\end{figure}

A key application in exact linear algebra is the resolution of
triangular systems over finite fields presented in Figure~\ref{trsm}.
A delayed prime field arithmetic version of this algorithm can be
constructed by simply doing a delayed matrix multiplication on the
operation \texttt{$B_1:= B_1 - A_2X_2$}. Listing~\ref{lst/mm} performs
such matrix multiplication \texttt{DGEMM\_NEG} with no reduction. We
used naming conventions of BLAS and LAPack for the function and the
parameter names. For the sake of simplicity some parameters have be
omitted and some function names were slightly modified.

The \texttt{DREMM} function computes the remainder modulo $p$ of all
the components of a matrix. We are overspecifying it for the purpose
of this presentation as our proof never use the exact value of these
remainders but the sole property that they are between $0$ and $p-1$.
In the definition of \texttt{is\_exact\_int\_mat}, we use the
predicate \texttt{exists}. The property described is that $X[i \times
LDX+j]$ is not any float, but it is an integer. To describe this, we
require (or prove) that there exists an integer equal to the
floating-point value of $X[i \times LDX+j]$. We also use \verb+\old+
in the annotations: the value \verb+\old(X)+ represents the value of
the variable {\tt X} before the function execution. It allows us to
specify the outputs depending on the inputs, even if the pointed
values are modified. Some ghost variables are introduced in our
example to ease the proofs. A mechanism not presented here prevents
such ghost variables to remain in the code once compiled.

\begin{lstlisting}[float,caption=Matrix-matrix multiplication Y <- Y - AX and component-wise remainder,label=lst/mm]
#include <math.h>
typedef double fp;
#define FP_ESPILON DBL_EPSILON

/*@ logic real epsilon() { 2^^(-53) } @*/
/*@ logic real max_int() { 2 / epsilon() } @*/

/*@ predicate is_exact_int_mat (fp *X, int LDX, int N, int M) {
  @   \valid_range(X,0,LDX*N) && M <= LDX &&
  @   \forall int i; \forall int j; 0 <= i < N && 0 <= j < M =>
  @      \round_error(X[i*LDX+j])==0 && \exists int v;  X[i*LDX+j]==v
  @ } @*/

/*@ predicate is_exact_int_mat_bounded_by
  @ (fp *X, int LDX, int N, int M, int min, int max) {
  @   is_exact_int_mat(X,LDX,N,M) && \forall int i; \forall int j;
  @     0 <= i < N && 0 <= j < M => min <= X[i*LDX+j] <= max
  @ } @*/

/*@ requires (p-1)*(p-1)*M <= max_int() && 
  @   is_exact_int_mat_bounded_by(Y,LDY,N,K,0,p-1) &&
  @   is_exact_int_mat_bounded_by(A,LDA,N,M,0,p-1) &&
  @   is_exact_int_mat_bounded_by(X,LDX,M,K,0,p-1)
  @ assigns Y[..]
  @ ensures 
  @   is_exact_int_mat_bounded_by(Y,LDY,N,K,(1-p)*(p-1)*M,p-1) @*/
void DGEMM_NEG (int N, int M, int K, int p,
                fp *A, int LDA, fp *X, int LDX, fp *Y, int LDY) {
  int i, j, k; fp oYiK;
  for (i = 0; i < N; i++)
    for (j = 0 ; j < M; j++)
      for (k = 0; k < K; k++) {
        oYik       = Y[i*LDY+k];
	Y[i*LDY+k] = Y[i*LDY+k] - A[i*LDA+j] * X[j*LDX+k];
        /*@ assert -1 <= Y[i * LDY + k] / ((p-1)*(p-1)*(j+1)) &&
                         Y[i * LDY + k] <= p-1                @*/
      }
}

/*@ requires is_exact_int_mat(X,LDX,N,K)
  @ assigns X[..]
  @ ensures is_exact_int_mat_bounded_by(X,LDX,N,K,0,p-1) &&
  @   \forall int i; \forall int j; 0 <= i < N && 0 <= j < K =>
  @     \exists int d; X[i*LDX+j] == \old(X[i*LDX+j]) + d*p @*/
void DREMM (int N, int K, int p, fp *X, int LDX) {
  int i, k;
  for (i = 0; i < N; i++) for (k = 0; k < K; k++) {
    X[i*LDX+k] = fmod (X[i*LDX+k], p);
    if (X[i*LDX+k] < 0) X[i*LDX+k] += p; 
  }
}
\end{lstlisting}

Let $M$ be the width of matrix $A_2$ or the height of vector $X_2$ and
{\tt FP\_EPSILON} be the machine $\epsilon$ for {\tt fp} in {\tt
  float}, {\tt double} or {\tt long double} and {\tt FP} in {\tt FLT},
{\tt DBL} or {\tt LDBL} respectively. The \texttt{DREMM} reduction can
be delayed until the end of \texttt{DGEMM\_NEG} provided
$$(p-1)^2\times M \le 2/\mathtt{FP\_EPSILON},$$
as all the number between 0 and $2/\mathtt{FP\_EPSILON}$ can be
represented exactly with type {\tt fp}.

Assertions on the \verb+DGEMM_NEG+ function generate proof obligation
$$
-1 \le \frac{ Y[i \times LDY + k]                                                 }{(p - 1)^2 \times (j + 1)}
     = \frac{oY i              k  - A[i \times LDA + j] \times X[j \times LDX + k]}{(p - 1)^2 \times (j + 1)}
$$
for all iterations defined by $i$, $j$ and $k$. It can be proved by
induction on $j$ with the following lemma that we proved in Coq and
similar ones for the other relations.
$$
\forall ~~ a, b, c, d, e \in \R ~~~ ; ~~~ e \le \frac{a}{b} ~~ \wedge ~~
                                          e \le \frac{c}{d} ~~ \wedge ~~
                                          0 <   b \times d
                ~~~ \Longrightarrow  ~~~  e \le \frac{a+c}{b+d}
$$

Gappa does not handle arrays so we have to rename $A[i \times LDA +
j]$ to {\tt Aik}, $X[j \times LDX + k]$ to {\tt Xjk} and $Y[i \times
LDY+k]$ to {\tt Yik}.  The following Gappa text should be sufficient
to prove the generic case of the induction if our lemma is added to
Gappa.

\begin{lstlisting}
{
  p in [2,1b53]                                ->
  j in [1,1b53]                                ->
  oYik in [-1b53,1b53]                         ->
  Aij/(p-1) in [0,1]                           ->
  Xjk/(p-1) in [0,1]                           ->
  oYik / ((p-1)*(p-1)*j) >= -1                 ->
  ((p-1)*(p-1)*j) * ((p-1)*(p-1)) >= 0         /\
  -Aij*Xjk / ((p-1)*(p-1)) >= -1               /\
  (oYik - Aij*Xjk) / ((p-1)*(p-1)*(j+1)) >= -1 
}
-Aij*Xjk / ((p-1)*(p-1)) ->
  - (Aij/(p-1)) * (Xjk/(p-1)) { (p-1) <> 0 };
(oYik - Aij*Xjk) / ((p-1)*(p-1)*(j+1)) ->
  (oYik + (- Aij*Xjk)) / (((p-1)*(p-1)*j) + ((p-1)*(p-1)))
  { ((p-1)*(p-1)*(j+1)) <> 0 , ((p-1)*(p-1)*j) * ((p-1)*(p-1)) <> 0 };
\end{lstlisting}

A Gappa file usually starts with aliases.  Gappa uses them for its
outputs and in the formal proof instead of machine generated names.
We do not need any alias here.  Identifiers are assumed to be
universally quantified over the set of real numbers the first time
Gappa encounters them.

The main assertion is written between brackets~({\tt \{ \}}). The
hypotheses end with the last line finished by an implication sign
({\tt ->}).  Each states that a variable or an expression is within an
interval or bounded. Note that $p1 \rightarrow p2 \rightarrow p3$ is
logically equivalent to $p1 \wedge p2 \rightarrow p3$.  The goal is
next.  It is a conjunction (\verb+/\+).  Statements about intermediate
variables are given as goals to force Gappa to establish them first.

The statement after the brackets are hints that propose replacements.
Gappa replaces the left side of the {\tt ->} sign by the right side as
soon as it encounters the former. It also tries to prove that the
replacements are valid. They help Gappa identify the proper theorems
syntactically. Conditions on the validity of the hints are expressed
between brackets.

\section{Variable interchange in a predicate}
\label{sec/xch}

\begin{lstlisting}[float,caption=Delayed solution of a unitary triangular system over a finite field,label=lst/lz]
/*@ predicate delay(int N, int p) @*/
/*@ logic int l_pmax(int n) @*/

// Floating point exact solution to a small unitary triangular system
/*@ requires N <= 54 &&  
  @   is_exact_int_mat_bounded_by(X,LDX,N,K,0,l_pmax(N)-1) &&
  @   is_exact_int_mat_bounded_by(A,LDA,N,N,0,l_pmax(N)-1)
  @ assigns X[..]
  @ ensures
  @   is_exact_int_mat_bounded_by(X,LDX,N,K,-max_int(),max_int()) @*/
void DTRSM (int N, int K, fp *A, int LDA, fp *X, int LDX) {
  int i, j, k;
  for (i = N-2; i >= 0; i--)
    for (j = i+1; j < N; j++)
      for (k = 0 ; k < K; k++)
        X[i*LDX+k] = X[i*LDX+k] - A[i*LDA+j] * X[j*LDX+k];
}

/*@ requires (p-1)*(p-1)*N <= max_int() &&
  @   is_exact_int_mat(A,LDA,N,N) && is_exact_int_mat(B,LDB,N,K)
  @ assigns B[..]
  @ ensures is_exact_int_mat(B,LDB,N,K) @*/
void LZ_TRSM (int N, int K, int Nmax, int p,
	      fp *A, int LDA, fp *B, int LDB) {
  if (N <= Nmax) {
    /*@ assert N <= 54 && delay(N,p) @*/
    DTRSM (N, K, A, LDA, B, LDB); DREMM (N-1, K, p, B, LDB);
  } else {
    int P = N/2, G = N - P;
    LZ_TRSM (G, K, Nmax, p, A+P*(LDA+1), LDA, B+P*LDB, LDB);
    DGEMM_NEG (P, G, K, p, A+P, LDA, B+P*LDB, LDB, B, LDB);
    DREMM (P, K, p, B, LDB);
    LZ_TRSM (P, K, Nmax, p, A, LDA, B, LDB);
  }
}


/*@ ensures \forall int N; N <= \result => delay (N, p) @*/
int Nmax (int p) {
  fp pp = 1, p2 = 1; int N;
  for (N = 0; ((p-1)*(pp+p2))/2 < 2 / 2^53; N++)
    {pp *= p; p2 *= p-2;};
  return N;
}

/*@ ensures \result==l_pmax(N) &&
  @   \forall int p; p <= \result => delay (N, p) @*/
int pmax (int N) {
  int p; for (p = 1; N <= Nmax(p); p++);
  return p-1;
}
\end{lstlisting}


For the sake of completeness we recall an optimal bound on integer
coefficients growth during backward substitution.

\begin{cor}{\cite[corollary 3.3]{jgd:2004:issac}}
  Let $A\in\Z^{N \times N}$ be a unit diagonal upper triangular
  matrix, and $b\in\Z^N$, with $|A|,|B|\leq p-1$. Then $x\in\Z^N$ the
  solution of the system $Ax=B$ is such that
  $$ |x| \leq \frac{p-1}{2}\left[ p^{N-1} + (p-2)^{N-1}\right],$$%
  and this bound is optimal.
\end{cor}

As this formula also bounds all the intermediate values, enough bits
are available to combine a few recursion steps of \texttt{LZ\_TRSM}
without reduction and guarantee that all numerical results are exact.
We present in Listing~\ref{lst/lz} an improved version for the delayed
prime field arithmetic by replacing the last levels of the recursion
by calls to \texttt{DTRSM} numerical solver according to the above
corollary.  Recursion is stopped for the maximal integer $N_{\max}$
such that
\begin{equation}
\frac{p-1}{2}\left[ p^{N_{max}-1} + (p-2)^{N_{max}-1}\right] \le 2/\mathtt{FP\_EPSILON}.
\end{equation}


Function {\tt Nmax} of Listing~\ref{lst/lz} uses a strict inequality
in equation (1) to control the loop because it computes the bound with
the same floating point format than the one used by \texttt{LZ\_TRSM}.
We focus now on proving that the \texttt{DTRSM} function invoked by
\texttt{LZ\_TRSM} never produce any round-off error.

One could port the proof of Corollary 3.3~\cite{jgd:2004:issac} to Coq
to finish the proof of correction. We decided to use Gappa and simple
techniques that could be made automatic. Syntactically, the
\verb+DTRSM+ function is invoked by \verb+LZ_TRSM+ only if the
condition $delay(N,p)$ is fulfilled. We may define it by:

$$delay(N,p) = N \le N_{\max}(p).$$

Gappa does not handle loops and branches. We perform a case analysis
but $p$ may vary from $2$ to $94,906,266$. On the other hand, {\tt N}
varies only between $2$ and $54$. Table~\ref{tab/pmax} presents the
value computed by the \verb+pmax+ function.  Variable interchange
should allow to prove that \verb+DTRSM+ is invoked only on the
condition
$$delay(N,p) \Longleftrightarrow p \le p_{\max}(N).$$

\begin{table}[t]
  \caption{Maximum value {\tt pmax} of parameter {\tt p} for each value of parameter {\tt N} allowed}
  \label{tab/pmax}
  $$
  \begin{array}{|c|c|c|c|c|c|c|c|c|c|c|c|} \hline
    N         & 2        & 3      & 4    & 5    & 6   & 7   & 8      & 9      & ~ 10 ~ & ~ 11 ~ & ~ 12 ~ \\ \hline
    p_{\max}  & 94906266 & 208064 & 9739 & 1553 & 457 & 191 & ~ 97 ~ & ~ 59 ~ & 39     & 29     & 19     \\ \hline
  \end{array}
  $$
  $$
  \begin{array}{|c|c|c|c|c|c|c|c|} \hline
    N        & ~ 13 ~ & ~ 14 ~ & ~ 15 ~ & ~ 16\cdots19 ~ & ~ 20\cdots23 ~ & ~ 24\cdots34 ~ & ~ 35\cdots54 ~ \\ \hline
    p_{\max} & 17     & 13     & 11     & 7              & 5              & 3              & 2              \\ \hline
  \end{array}
  $$
\end{table}

\begin{table}[t]
  \caption{Time to establish that no round-off occurred and generate a Coq proof script}
  \label{tab/time}
  $$
  \begin{array}{|c|c|c|c|c|c|c|c|c|c|c|c|} \hline
    \text{Script} & N               & 2     & 3     & 4     & 5     & 6     & 7     & 8     & 9     & 10    & 11    \\ \hline
    \text{no}     & \text{time} (s) & 00.02 & 00.06 & 00.10 & 00.19 & 00.29 & 00.43 & 00.57 & 00.74 & 00.93 & 01.15 \\ \cline{1-1}\cline{3-12}
    \text{Coq}    &                 & 00.02 & 00.06 & 00.13 & 00.22 & 00.35 & 00.50 & 00.87 & 01.35 & 01.84 & 02.43 \\ \hline
  \end{array}
  $$
  $$
  \begin{array}{|c|c|c|c|c|c|c|c|c|c|c|c|} \hline
    \text{Script} & N               & 12    & 13    & 14    & 15    & 17    & 19    & 23    & 34     & 54    \\ \hline
    \text{no}     & \text{time} (s) & 01.38 & 01.63 & 01.91 & 02.90 & 02.90 & 03.73 & 05.93 & 13.71  & 38.32 \\ \cline{1-1}\cline{3-11}
    \text{Coq}    &                 & 03.22 & 03.77 & 05.26 & 07.13 & 12.22 & 19.50 & 38.13 & 421.4  & 5767  \\ \hline
  \end{array}
  $$
\end{table}

A C++ class produces a trace for {\tt N} between 2 and 54 where all
branches and control statements have been removed. Each trace contains
only floating point operations.  Table~\ref{tab/time} gives the time
needed by Gappa on a 1.86~GHz Quad-Core Intel Xeon Processor with
2x4MB L2 cache and 2x1GB of memory, to produce each of the proofs that
no round-off error occurred for all the values of $N$ between $2$ and
$54$. That ends the proof of correction of the algorithm.

\section{Perspectives and concluding remarks}
\label{sec/conc}

This report presents a new use of Gappa based on proof obligations
generated from a trace of execution~\cite{MelPio07}.  It would enable
us to prove in Coq in the future that expression swell within the
studied algorithm of delayed finite field arithmetic does not
introduce round off errors \cite{jgd:2004:issac}. This full
certification will not be obtained by porting the initial proof to Coq
but by a case analysis on the 53 possible values of one argument $N$.
Pieces of the formal proof have been generated by Gappa for each
individual value of $N$.


Our approach can be easily reproduced to other exact linear
applications over finite fields.  More precisely, the FFLAS-FFPACK
project has been successful on using delayed prime field arithmetic
for linear algebra applications.
\begin{center}
  \url{http://ljk.imag.fr/membres/Jean-Guillaume.Dumas/FFLAS/}
\end{center}



\bibliographystyle{abbrv}
\bibliography{daumas_ref,alternate}

\end{document}